\documentclass[usenatbib]{mn2e}

\usepackage{graphicx,tabularx,amsmath,amssymb,upgreek,wasysym,mathtools,multirow}
\numberwithin{equation}{section}
\title[Dust in the Eye of Andromeda]{Dust in the Eye of Andromeda}
\author[K. A. Marsh, A. P. Whitworth, M. W. L. Smith, O. Lomax, and S. A. Eales]
{K. A. Marsh\thanks{E-mail: Ken.Marsh@astro.cf.ac.uk}, 
A. P. Whitworth, M. W. L. Smith, O. Lomax, and S. A. Eales \\ \\
School of Physics and Astronomy, Cardiff University, Cardiff CF24 3AA, Wales, UK}
\begin{document}
\maketitle
\label{firstpage}

%%%%%%%parameters with
\begin{abstract} 
We present new {\it Herschel\/}-derived images of warm dust in the Andromeda 
Galaxy, M31, with unprecedented spatial resolution ($\sim30$ pc), 
column density accuracy, and constraints on the three-dimensional 
distributions of dust temperature and dust opacity index (hence
grain size and composition), based on the new {\sc ppmap} Bayesian analysis 
procedure. We confirm the overall radial variation of dust opacity index 
reported by other recent studies, including the central decrease 
within $\sim3$ kpc of the nucleus. We also investigate the detailed
distribution of dust in the nuclear region,
a prominent feature of which is a $\sim500$ pc bar-like structure
seen previously in H$\alpha$. The nature of this feature has been the subject 
of some debate. Our maps show it to be the site of the warmest dust, with a
mean line-of-sight temperature $\sim30$ K.
A comparison with the stellar distribution, based on 2MASS data, provides
strong evidence that it is a gravitationally induced bar. A comparison
with radial velocity maps
suggests the presence of an inflow towards the nucleus from opposite directions
along this bar, fed presumably by the nuclear spiral
with which it appears to connect. Such behaviour is common in large-scale
bars in spiral galaxies, as is the phenomenon of nested bars
whereby a subkiloparsec nuclear bar
exists within a large-scale primary bar. We suggest that M31 represents
an example of such nesting.
\end{abstract}
%%%%%%%


\section{Introduction}

High resolution mapping of dust in spiral galaxies can provide information
regarding star formation rates, the interstellar radiation field,
and the global dynamics of the galaxies themselves. Such studies have 
been greatly aided by data from the {\it Herschel\/} Space Observatory whose
wavelength coverage (70--500 $\mu$m) includes the
peak of the spectral range of cool dust emission. In terms of spatial
resolution, the most favourable candidate for study is
the Andromeda galaxy, M31, 
being our nearest spiral neighbor, at an estimated distance of 780 kpc 
\citep{stan98,vil06}. Knowledge of the distribution of cool dust provides
constraints on a number of physical phenomena in this galaxy and hence the
{\it Herschel\/} observations represent an important component in
the suite of multiwavelength data which extends from 
radio wavelengths to X-rays. 

The spatial structure of M31, as viewed in the far infrared, is dominated by 
a set of bright concentric rings whose approximate radii, in order of decreasing
prominence, are 10, 5, and 15 kpc \citep{habing84,gordon06,fritz12,smith2012,
kirk2015,lewis2015}.  The disk of the galaxy is
inclined at $77^\circ$ to the plane of the sky, with a tilt axis at
position angle $38^\circ$ \citep{fritz12}.  There is also an inner 
ring, of approximate radius 0.7 kpc, which is less inclined than the main 
structure \citep{melch11,melch13} and which may represent the aftermath of a 
collision with the M32 dwarf galaxy \citep{block06}. 
Studies of the inner isophotes at
optical and near-infrared wavelengths have provided evidence of a large-scale
bar of length $\sim10$ kpc (\citet{ath06,beat07,op16,diaz17}, although the
existence of such a bar has been questioned by \citet{melch11}.

Evidence for significant radial variations in dust properties has been obtained
by \citet{smith2012} using fits of model spectral energy distributions (SEDs)
to {\it Herschel\/} continuum data at six bands in the wavelength range 
70--500 $\mu$m. The angular spatial resolution was $36''$, corresponding to a
linear resolution of $\sim140$ pc at the distance of M31.
They found that the dust opacity index, $\beta$, increases
inwards, reaching a maximum value at a radial distance $\sim 3$ kpc,
inside of which it then decreases towards the nucleus. A similar conclusion
has been reached by \citet{draine2014}.

In the present paper we present the results of our investigation of 
the dust distribution in Andromeda using the same
{\it Herschel\/} data as used by \citet{smith2012}, but processed
using a new Bayesian analysis procedure, {\sc ppmap} 
\citep{mar15} which
yields significantly higher spatial resolution (30 pc).
This procedure, outlined in the next section, overcomes a number 
of limitations in standard analysis techniques. These include the requirement 
that all input images be degraded 
to the lowest resolution (normally $36''$ in the case of {\it Herschel\/}
data) and the assumption that the dust has uniform temperature and composition 
along each line of sight. Such techniques return typically a pair of 
two-dimensional maps, one giving a {\it representative} 
dust temperature, ${\hat T}_{_{\rm D}}(x,y)$, and the other an approximate 
line-of-sight column density, ${\hat N}(x,y)$.
Here $N$ is the total column density of material (dust plus gas), expressed 
in units of hydrogen molecule masses
per unit area at the source, and the hat symbols (e.g. ${\hat N}$) 
denote values obtained with the standard analysis procedure.

Expressing the dust distribution in this way, i.e., in terms of column density, 
involves some grain model assumptions. Specifically,
we need to assume a value for the opacity per unit mass at 
some reference wavelength which we take as 300 $\mu$m, as discussed in the 
next section. However, {\sc ppmap}\ can alternatively
produce maps of {\it optical depth\/} at that wavelength, 
free from any grain model assumptions. The results
could then be compared with optical and/or near-infrared extinction measurements
to extend the extinction curve to 300 $\mu$m in a model-independent way.
Such an analysis forms the basis of a companion paper (Whitworth \& Marsh,
in preparation).

\begin{figure*}
\includegraphics[width=150mm]{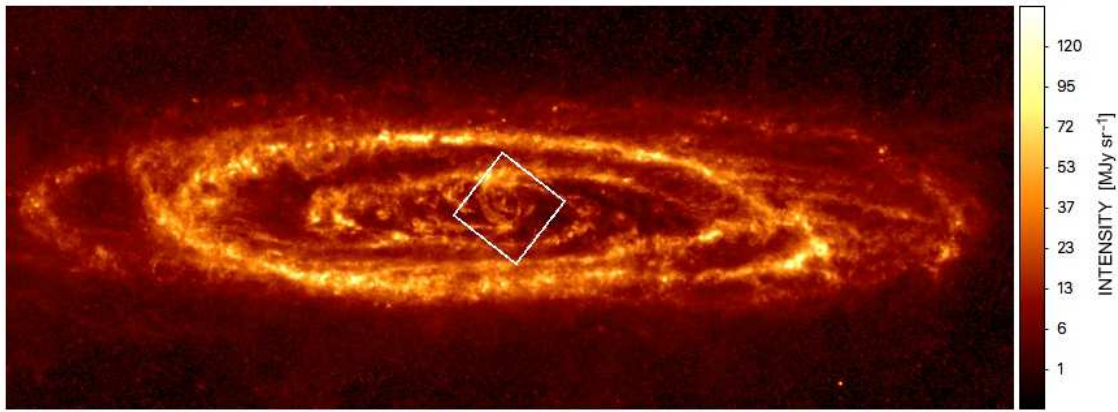}
\caption{M31 as viewed in the 250 $\mu$m band by the {\it Herschel\/}
SPIRE instrument.  The square box 
delineates the ``ZoomZone" discussed in the text. Its size is
700 arcsec on a side, corresponding to 2.65 kpc based on an
assumed distance of 780 kpc.}
\label{fig1}
\end{figure*}

\section{Methodology}

In contrast with previous approaches to the dust mapping problem,
the {\sc ppmap} procedure drops the assumptions of uniform 
dust temperature and opacity index along the line of sight
\citep{mar15}\footnote{The original version treated the opacity index,
$\beta_{_{\rm D}}$, as constant. We have extended the algorithm by
treating $\beta_{_{\rm D}}$ as an additional state variable. Further details
are given in Appendix A.}.
After taking proper account of the point-spread function and spectral response of the telescope, it returns a pair of four-dimensional hypercubes containing, respectively, the expectation values of differential column density,
\begin{eqnarray}\nonumber
N_{_{T\beta}}&\equiv&\frac{\partial^2\!N}{\partial T_{_{\rm D}}\,\partial\beta_{_{\rm D}}}\,,
\end{eqnarray}
{\it and\/} their associated uncertainties, as a function of $(x,y,T_{_{\rm D}},\beta_{_{\rm D}})$.
Here $\beta_{_{\rm D}}$ is the opacity index of the dust in the far-infrared 
i.e. 
\begin{eqnarray}
\beta_{_{\rm D}}&=&-\,\frac{d\log\kappa_\lambda}{d\log\lambda}\,;
\end{eqnarray}
$\kappa_\lambda$ is the mass opacity coefficient of the dust at wavelength 
$\lambda$, and the derivative is evaluated in the wavelength interval 
covered by {\it Herschel\/}. Thus $N_{_{T\beta}}
(x,y,T_{_{\rm D}},\beta_{_{\rm D}})$ is the best estimate of the line-of-sight
column density at $(x,y)$, in unit 
dust temperature interval about $T_{_{\rm D}}$ and unit 
opacity index interval about $\beta_{_{\rm D}}$. 
The procedure assumes only that the emitting dust is in radiative equilibrium, and that, at the observed wavelengths,  the emission is optically thin. 
It works by defining a grid of discrete values of $x$, $y$, $T_{_{\rm D}}$ 
and $\beta_{_{\rm D}}$, which represent sampling locations in the continuous
state space of $(x,y,T_{_{\rm D}},\beta_{_{\rm D}})$. 
It then iterates towards the set of differential column densities, in the
vicinities of those grid points, that best reproduces the observed 
monochromatic intensities.
The resolution on the plane of the sky $(x,y)$ is increased by a factor 
of about 4.5 relative to the standard analysis procedure, yielding maps of
angular resolution of $\sim8''$, sampled by $4''$ square pixels 

The dust temperature sampling interval is arbitrary. For this work, we have used twelve representative dust temperatures that are equally spaced in logarithm between $\,10.0\,{\rm K}\,$ and $\,50.0\,{\rm K}$, i.e. $T_1\!=\!10.0\,{\rm K}$, $T_2\!=\!11.6\,{\rm K}$, $T_3\!=\!13.4\,{\rm K}$, $T_4\!=\!15.5\,{\rm K}$, $T_5\!=\!18.0\,{\rm K}$, $T_6\!=\!20.8\,{\rm K}$, $T_7\!=\!24.1\,{\rm K}$, $T_8\!=\!27.8\,{\rm K}$, $T_9\!=\!32.2\,{\rm K}$, $T_{10}\!=\!37.3\,{\rm K}$, $T_{11}\!=\!43.2\,{\rm K}\,$ and $T_{12}\!=\!50.0\,{\rm K}$. 

The opacity index sampling interval is also arbitrary, and we have used 
four representative values equally spaced linearly 
between $\,\beta_{_{\rm D}}\!=\!1.5\,$ and $\,\beta_{_{\rm D}}\!=\!3.0\,$, 
i.e. $\,\beta_1\!=\!1.5$, $\,\beta_2\!=\!2.0$, $\,\beta_3\!=\!2.5\,$ 
and $\,\beta_4\!=\!3.0$. If we were to invoke more representative 
dust temperatures and/or opacity indices, the computational cost of the 
analysis would increase proportionately but further accuracy would not 
necessarily be gained.

Our column density
scale is based on an assumed opacity of 0.1 cm$^2$ g$^{-1}$ at a wavelength
of 300 $\mu$m. The reference opacity is
defined with respect to total mass (dust plus gas). Although observationally
determined it is consistent with a gas to dust ratio of 100 \citep{hil83},
and we use it here on the assumption that interstellar dust grains in 
M31 have similar properties to those in our own Galaxy.

Given the $\,N_{_{T\beta}}\,$ hypercube, we can integrate out $\beta$ to obtain two three-dimensional data-cubes containing, respectively, representative values of 
\begin{eqnarray}
N_{_T}&\equiv&\frac{dN}{dT_{_{\rm D}}}\;\,=\;\,\int\limits_{{\rm all}\;\beta_{_{\rm D}}}\;\frac{\partial^2\!N}{\partial T_{_{\rm D}}\,\partial\beta_{_{\rm D}}}\;d\beta_{_{\rm D}}
\end{eqnarray}
and their associated uncertainties. $\,N_{_T}(x,y,T_{_{\rm D}})\,$ is the best estimate of column density along the line of sight at $(x,y)$, in unit dust temperature interval about $\,\sim\!T_{_{\rm D}}$. This data cube can be interpreted in two interesting ways. First, we can take slices at the different representative dust temperatures, $T_i$, and plot separately the column density of dust in each dust temperature interval, $((T_{i-1}T_i)^{1/2}\leq T_{_{\rm D}}\leq (T_iT_{i+1})^{1/2})$. Second, we can consider individual pixels separately, and analyse the mass-weighted distribution of dust temperature along the line of sight through the pixel. Since the dust temperature depends on the ambient radiant energy density, $U_{_{\rm RAD}}$ (in relatively unshielded regions, $T_{_{\rm D}}\propto U_{_{\rm RAD}}^{0.20\pm0.05}$), this can help to resolve regions along the line of sight that have different ambient radiation fields.

Similarly, we can integrate out $T_{_{\rm D}}$ to obtain two three-dimensional data-cubes containing, respectively, representative values of 
\begin{eqnarray}
N_{_\beta}&\equiv&\frac{dN}{d\beta_{_{\rm D}}}\;\,=\;\,\int\limits_{{\rm all}\;T_{_{\rm D}}}\;\frac{\partial^2\!N}{\partial T_{_{\rm D}}\,\partial\beta_{_{\rm D}}}\;dT_{_{\rm D}}
\end{eqnarray}
and their associated uncertainties. $\,N_{_\beta}(x,y,\beta_{_{\rm D}})\,$ is the best estimate of the line-of-sight column density at $(x,y)$, in unit opacity index interval about $\,\sim\!\beta_{_{\rm D}}$. Again, this data cube can be interpreted in two interesting ways. First, we can take slices at the different representative opacity indices, $\beta_j$, and plot the column density of dust in each opacity index interval, $(\beta_j-0.25\leq \beta_{_{\rm D}}\leq \beta_j+0.25)$. Second, we can consider individual pixels separately, and analyse the mass-weighted distribution of opacity index along the line of sight through the pixel. Since the opacity index depends on the constitution of the dust (size, composition, shape), this can help to distinguish regions along the line of sight that have different chemical histories. 

Next, we can integrate out both $\beta_{_{\rm D}}$ and $T_{_{\rm D}}$, to obtain two two-dimensional maps of
\begin{eqnarray}
N&=&\int\limits_{{\rm all}\;T_{_{\rm D}}}\;\frac{dN}{dT_{_{\rm D}}}\;dT_{_{\rm D}}
\end{eqnarray}
and its associated uncertainty. Here $N(x,y)$ gives the best estimate for the 
line-of-sight column density at $(x,y)$.

We can obtain an estimate of the mass-weighted mean dust temperature, ${\bar T}_{_{\rm D}}(x,y)$, along the line of sight at $(x,y)$,
\begin{eqnarray}
{\bar T}_{_{\rm D}}&=&\frac{1}{N}\,\int\limits_{{\rm all}\;T_{_{\rm D}}}\;T_{_{\rm D}}\;\frac{dN}{dT_{_{\rm D}}}\;dT_{_{\rm D}}\,,
\label{eqT}
\end{eqnarray}
and higher-order moments of the dust temperature distribution along the line of sight (standard deviation, skewness and kurtosis), as discussed below.

Finally, we can obtain an estimate of the mass-weighted mean opacity index, ${\bar\beta}_{_{\rm D}}(x,y)$, along the line of sight at $(x,y)$
using an integral analogous to Eq. (\ref{eqT}).

\section{Observations}

M31 was observed using the {\it Herschel\/} PACS and SPIRE instruments 
which provided continuum images in bands centred on
wavelengths of 70, 100, 160, 250, 350, and 500 $\mu$m. 
Full details of the observing strategy, calibration, and map-making 
procedures are given by \citet{fritz12} and \citet{smith2012}. As an example,
Fig.\ref{fig1} shows a SPIRE image of M31 in the 250 $\mu$m band.

The spatial resolution values of the scan maps, i.e., the beam sizes at full 
width half maximum (FWHM), are approximately
8.5, 12.5, 13.3, 18.2, 24.5, and 36.0 arcsec at the six wavelengths, 
respectively.  For our analysis we use PSFs
based on the measured {\it Herschel\/} beam profiles \citep{pog10,griffin13}. 

\section{Results}

The above data have been used to generate a 4D hypercube of differential
column density as a function of sky location, dust temperature, and
dust opacity index, from which we can derive various line-of-sight
integrated quantities such as the
integrated column density of dust plus gas, $N$, the mean 
dust temperature, ${\bar T_{_{\rm D}}}$, and the mean dust 
opacity index ${\bar \beta}_{_{\rm D}}$. 
Here, and in the following 
discussion, an overscore (e.g. ${\bar T}_{_{\rm D}}$ 
and ${\bar \beta}_{_{\rm D}}$) denotes mass-weighted averages along the 
line of sight.

\begin{figure}
\hspace*{-0.3cm}\includegraphics[width=90mm]{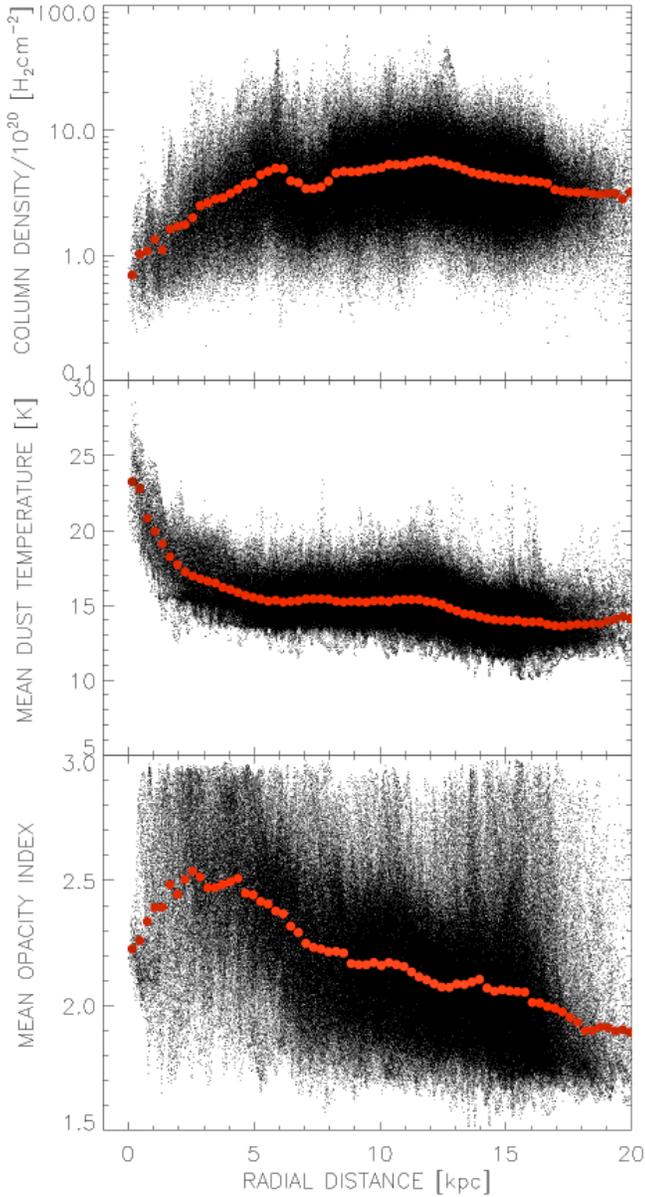}
\caption{Radial variation of total line-of-sight column density, 
$<\!\!\!N\!\!\!>$, mean dust temperature, 
$<\!\!\!{\bar T}_{_{\rm D}}\!\!\!>$, and mean dust opacity index,
$<\!\!{\bar\beta}_{_{\rm D}}\!\!>$. 
The red filled circles represent averages in 300 pc bins.}
\label{fig2}
\end{figure}

Fig. \ref{fig2} shows the radial variation of azimuthally averaged values of
these quantities, $<\!\!\!N\!\!\!>$, $<\!\!\!{\bar T}_{_{\rm D}}\!\!\!>$, 
and $<\!\!{\bar\beta}_{_{\rm D}}\!\!>$, at each
galactocentric radius. Galactocentric radii are estimated 
assuming that the midplane of Andromeda makes an angle of $\theta =77^{\rm o}$ 
with the plane of the sky, and that the position angle of the tilt axis 
is $38^\circ$.  We note that the radial plot of 
$<\!\!{\bar\beta}_{_{\rm D}}\!\!>$ in this figure indicates
a central dip in the dust opacity index, consistent with previous reports by
\citet{smith2012} and \citet{draine2014}. In particular there is detailed
correspondence with Fig. 13 of \citet{draine2014}, including the local
maximum at $r\sim3$ kpc.

\subsection{The central region: ``ZoomZone" }

\begin{figure}
\hspace*{-0.3cm}\includegraphics[width=90mm]{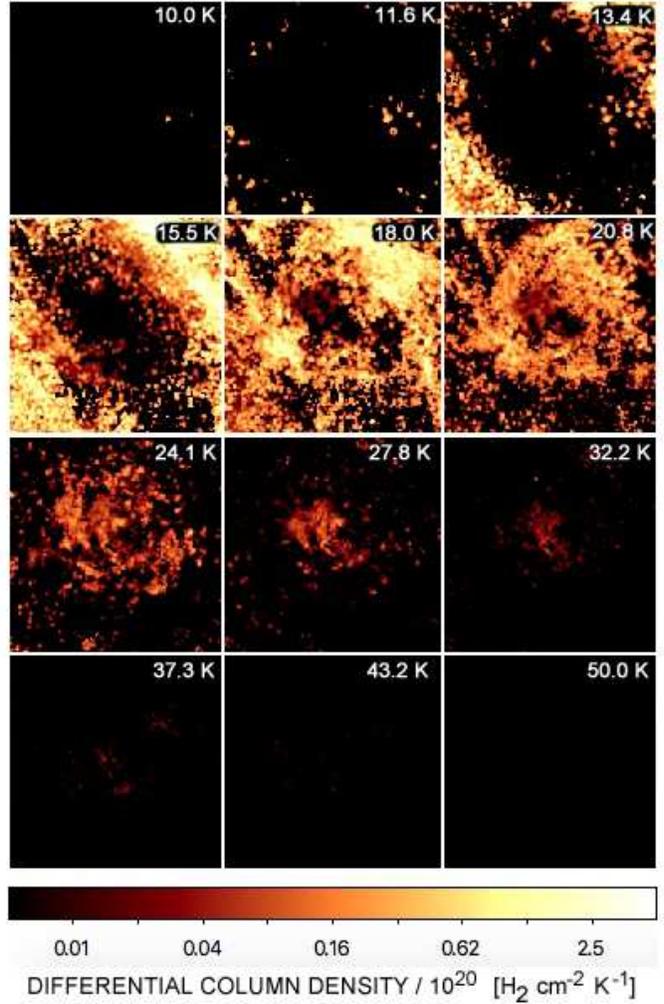}
\caption{Maps of differential column density as a function of temperature,
for the ZoomZone. These maps have been truncated at the $5\sigma$ level.
The field of view in each case is 700 arcsec (2.65 kpc)
square.}
\label{fig3}
\end{figure}

In the remainder of the paper, we focus on the small region of Andromeda that 
is marked with a square of width 700 arcsec (2.65 kpc) on Fig. 1, and refer to it as the ZoomZone.
It is chosen because it includes the circumnuclear environment, the site
of some interesting physical processes which are not well understood
\citep[see, for example ][]{li09}.
Fig. \ref{fig3} shows maps of temperature-differential column density, $N_T$, 
in this region, at all 12 temperatures. These maps, and the associated
uncertainty maps, are
available in the on-line material as described in Appendix B.

\begin{figure*}
\includegraphics[width=150mm]{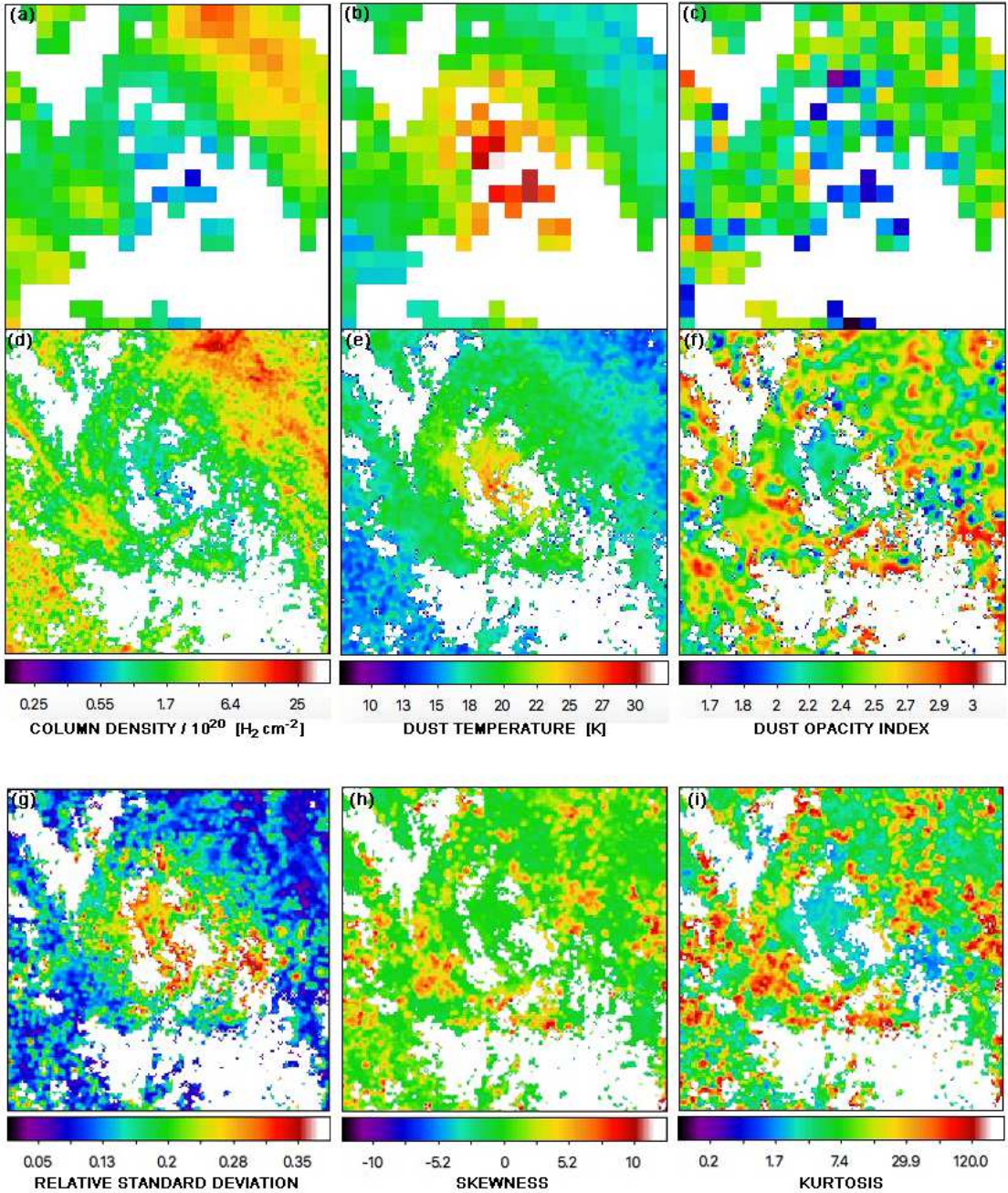}
\caption{Maps of line-of-sight integrated quantities for the ZoomZone,
each with a field of view of 700 arcsec (2.65 kpc) square, and each
truncated at the $5\sigma$ level with respect to column density.
{\it Top row:\/} Maps obtained by \citet{smith2012} using the standard
analysis procedure, viz. {\it (a)\/} Integrated column density, ${\hat N}$, 
{\it (b)\/}  mean dust temperature,
${\hat T}_{_{\rm D}}$ and {\it (c)\/} mean dust opacity index, 
${\hat \beta}_{_{\rm D}}$.
{\it Middle row:\/} The 
corresponding maps obtained with {\sc ppmap}, viz. {\it (d)\/} the integrated 
column density, $N$, {\it (e)\/} the mass-weighted mean dust temperature, 
${\bar T}_{_{\rm D}}$, and {\it (f)\/} the mass-weighted mean opacity index, 
${\bar\beta}_{_{\rm D}}$. 
{\it Bottom row:\/} Maps of the higher order moments of the dust
temperature distribution, viz. 
{\it (g)\/} the standard deviation, $\sigma_T$, expressed as a fraction
of the local mean temperature, ${\bar T}_{_{\rm D}}$,
{\it (h)\/} the skewness, and {\it (i)\/} the kurtosis.}
\label{fig4}
\end{figure*}

On the first  (top) row of Fig. \ref{fig4} are maps of (a) ${\hat N}$, (b) 
${\hat T}_{_{\rm D}}$ and (c) ${\hat \beta}_{_{\rm D}}$ obtained by 
\citet{smith2012} using the standard analysis procedure, cropped at
the boundaries of the ZoomZone. The second (middle) row shows the 
corresponding maps obtained with {\sc ppmap}, viz. (d) the integrated 
column density, $N$, (e) the mass-weighted mean dust temperature, 
${\bar T}_{_{\rm D}}$, and (f) the mass-weighted mean opacity index, 
${\bar\beta}_{_{\rm D}}$. These maps illustrate the greatly increased 
resolution obtained with {\sc ppmap}. In addition, they demonstrate the 
ability of {\sc ppmap} to dig out large quantities of cold dust that is 
underestimated by the standard analysis procedure. This is illustrated by 
the presence of lower mean dust temperatures (by up to 2 K) in panel (e)
than panel (b), while Fig. \ref{fig3} shows the presence of significantly
cooler dust ($\sim10$ K as opposed to $\sim15$ K) than indicated by the
standard analysis in the corresponding region.  As a consequence, in 
many areas {\sc ppmap} obtains a lower mean dust temperature 
(${\bar T}_{_{\rm D}}<{\hat T}_{_{\rm D}}$), and a higher net column density 
($N>{\hat N}$).  The third (bottom) row shows (g) the relative standard 
deviation, $\sigma_T/\bar{T}_{\rm D}$, (h) 
the skewness, Skew$(T)$, and (i) the kurtosis, Kurt$(T)$  of the 
dust temperature distribution along each line of sight. 
These three quantities are defined by:
\begin{eqnarray}
\sigma_T\!&=&\!\left(
\frac{1}{N}\!\int\limits_{{\rm all}\;T_{_{\rm D}}}
\frac{dN}{dT_{_{\rm D}}} (T_{_{\rm D}}-{\bar T}_{_{\rm D}})^2 
\,\,dT_{_{\rm D}} \right)^{1/2} \\
{\rm Skew}(T)\!&=&\!\left(
\frac{1}{N}\!\int\limits_{{\rm all}\;T_{_{\rm D}}}
\frac{dN}{dT_{_{\rm D}}} (T_{_{\rm D}}-{\bar T}_{_{\rm D}})^3 
\,\,dT_{_{\rm D}} \right)/ \sigma_T^3 \\
{\rm Kurt}(T)\!&=&\!\left(
\frac{1}{N}\!\int\limits_{{\rm all}\;T_{_{\rm D}}}
\frac{dN}{dT_{_{\rm D}}} (T_{_{\rm D}}-{\bar T}_{_{\rm D}})^4
\,\,dT_{_{\rm D}} \right)/ \sigma_T^4
\end{eqnarray}

\begin{figure*}
\includegraphics[width=150mm]{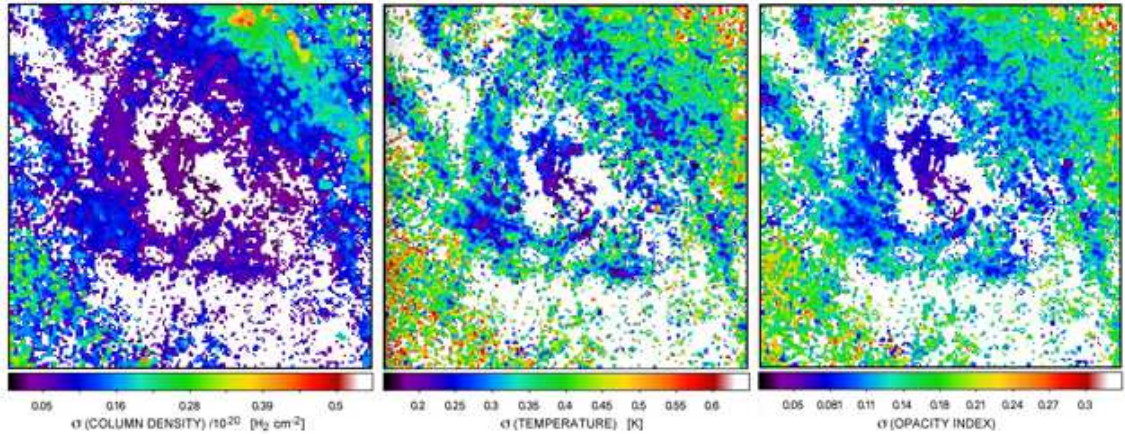}
\caption{Uncertainty maps of column density, temperature, and opacity
index for the ZoomZone,
each with a field of view of 700 arcsec (2.65 kpc) square. They represent the
uncertainty values corresonding to the maps in the middle row of
Fig. \ref{fig4}.}
\label{fig5}
\end{figure*}

They provide information on the form of the temperature 
distribution along each line of sight. In particular, for a pure Gaussian 
distribution we would expect a skewness of
zero and a kurtosis of 3. Examination of Fig. \ref{fig4} shows that
these conditions are met over much of the ZoomZone, although there are
localised regions in which the kurtosis is large ($\stackrel{>}{_\sim}100$
in many cases). Large values of the kurtosis indicate the presence of
a significant number of
high-temperature outliers in the tail of the distribution. It is not clear 
what high-kurtosis implies physically;
comparison with published positions of planetary nebulae \citep{merr06},
X-ray sources \citep{stiele2011}, and supernova remnants \citep{lee14}
indicate little or no correspondence with high-kurtosis features in the 
ZoomZone, while none of the HII regions listed by \citet{azi11} falls within 
that zone. However, comparison over the full field of M31 shows a 
high degree of correlation with the locations of HII regions from the 
\citet{azi11} survey, in the sense that HII regions are preferentially
located in regions of high temperature kurtosis and, interestingly, low
standard deviation of temperature. This suggests that high kurtosis might be 
a signpost of star formation, and this is confirmed by comparison with
a map of star formation rate \citep{ford13}. It would clearly be interesting
to investigate what kurtosis and the other temperature moments tell us
about star formation, but since the current focus is on the nuclear
region, we defer a discussion of more global phenomena to a future paper.

The uncertainty maps corresponding to the middle row of Fig. \ref{fig4}
(column density, temperature, and opacity index) are presented in
Fig. \ref{fig5}.

\begin{figure*}
\includegraphics[width=150mm]{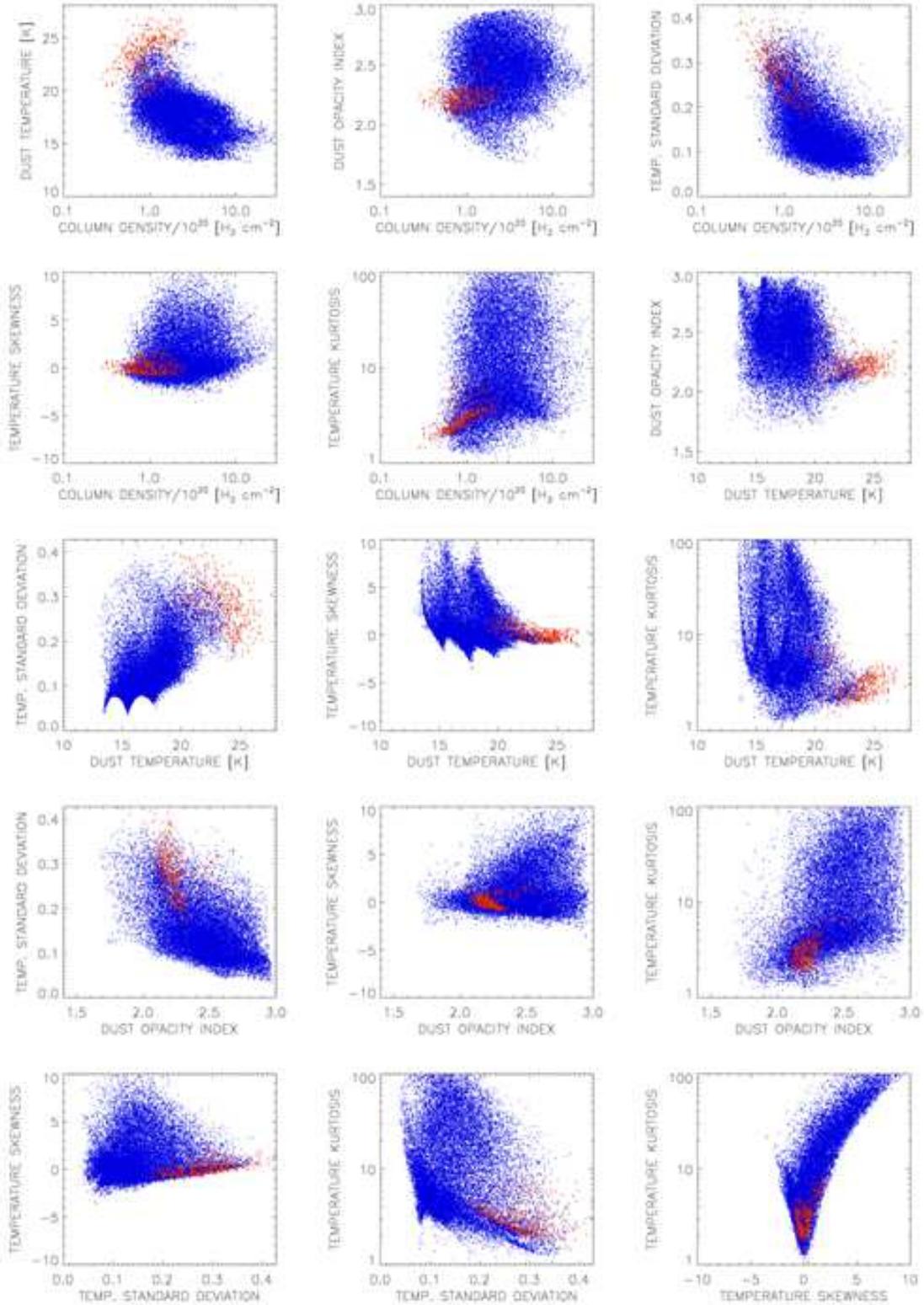}
\caption{Scatter plots of line-of-sight integrated quantities for the ZoomZone.
The standard deviation of temperature is expressed as a fraction of the
local mean line-of-sight temperature.
The red points represent spatial locations within the central bar-like
feature in the nuclear spiral, within 270 pc of the nucleus. The spiky
lower envelopes in the plots of $\sigma_T$ and Skew$(T)$ versus 
$T_{_{\rm D}}$ are due to the discretisation
of temperature values in {\sc ppmap}.}
\label{fig6}
\end{figure*}

Fig. \ref{fig6} shows a set of scatter plots, for the ZoomZone, between all 
pairs of the six quantities, i.e., column density, temperature, opacity 
index, and temperature standard deviation, skewness and kurtosis.
Various correlations are apparent, including an
anticorrelation between kurtosis and standard deviation which, as mentioned
above, appears to be a global characteristic. Also, particularly relevant here 
are the correlations involving the nuclear
bar-like feature, plotted in red. This feature clearly shows some
systematic behaviour, one example being that the dust opacity index,
${\bar\beta}_{_{\rm D}}(x,y)$,
is positively correlated with dust temperature in the interior of the
nuclear spiral (within 270 pc of the nucleus). This goes against the 
larger-scale trend in which ${\bar\beta}_{_{\rm D}}(x,y)$,
is {\it anticorrelated} with $\bar T_{_{\rm D}}$ within the
central $\sim3$ kpc. As discussed by \citet{smith2012}, various explanations are
possible for that anticorrelation, such as the destruction of grain
ice-mantles in the increasingly intense interstellar radiation field
towards the centre of the galaxy. However, the fact that it switches to
a positive correlation near the very centre would seem to argue against
that explanation, although it might instead indicate 
that a separate population of grains is present.

\begin{figure*}
\includegraphics[width=100mm]{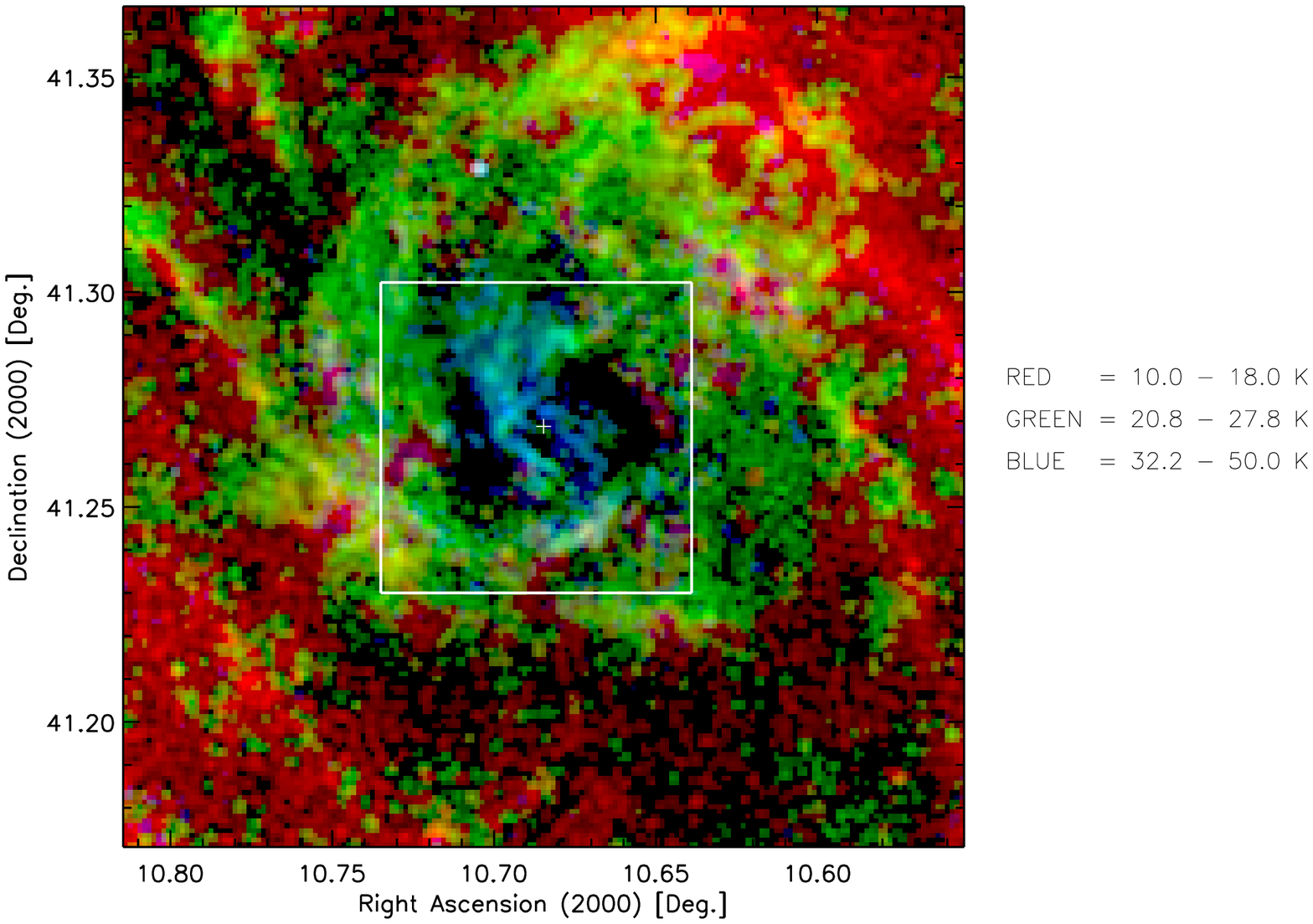}
\caption{Three-temperature composite of the ZoomZone, where
the dust temperature ranges correspond to the colours indicated. The
white square delineates the inner portion corresponding to the field of
view of the extinction map of \citet{dong16}.}
\label{fig7}
\end{figure*}

Fig. \ref{fig7} represents a three-temperature composite for the ZoomZone.
It shows that the innermost portion is dominated by a bar-like feature
which appears to consist of a pair of hot, roughly parallel, strands with a 
central crossover. It is distinctly visible at other wavelengths,
such as H$\alpha$ \citep{jac85} and 70 $\mu$m (in our {\it Herschel\/}
image), although less so at 24 $\mu$m \citep{gordon06}
due to confusion by a dense concentration of bulge stars. 
Its nature has been the subject of debate, but
with a length of only $\sim500$ pc, it clearly does not correspond to 
the much larger ($\sim10$ kpc) bar inferred from near-infrared morphology 
and dynamical simulations \citep{ath06}.

\begin{figure*}
\includegraphics[width=130mm]{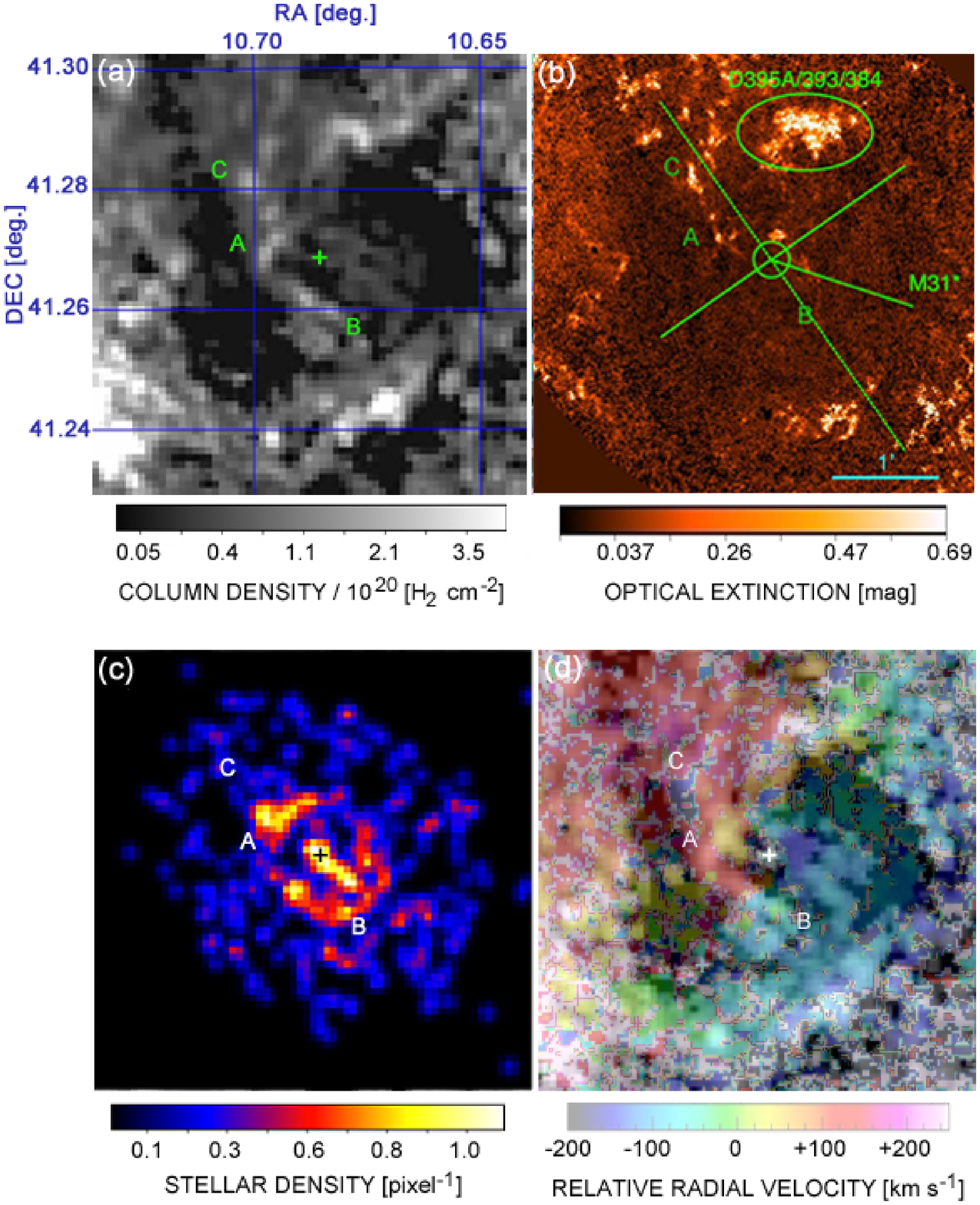}
\caption{Comparison of column density, extinction, stellar density, and
radial velocity
distributions for the central bulge region of M31, as delineated by the
white square in Fig. \ref{fig7}. {\it (a)\/} Total 
column density; {\it (b)\/} Optical extinction at a wavelength of 5447 \AA, 
from \citet{dong16}; {\it (c)\/} Stellar density in units of
stars per pixel, representing the number
density of  2MASS sources with $J\le14$, smoothed by a Gaussian
kernel of FWHM = 2 pixels (8 arcsec);
{\it (d)\/} Radial velocity with respect to the systemic value of
300 km s$^{-1}$, superposed on the total column density distribution.
The radial velocities are from [NII] observations published previously
by \citet{boule87}.
In panels (a), (c), and (d), the position of the nucleus is marked
by a plus sign (+). In (b) it is labelled as M31*. 
``A", ``B", and ``C" are reference points discussed in the text.}
\label{fig8}
\end{figure*}

\subsection{The nature of the bar-like feature}

Some information on the 3D geometry of the bar-like feature can be obtained
from a comparison of the column density distribution derived here
on the basis of dust emission, and a
previously published extinction map \citep{dong16}, as shown in panels
(a) and (b) of Fig. \ref{fig8}.
These maps show some distinct differences. For example
the filamentary structure which extends between reference points ``A" and ``B"
in the column density map is absent in the extinction map, suggesting that
it lies behind the bulge and hence behind the source of illumination which
enables foreground features to be detected in extinction. This behaviour was
also noted by \citet{dong16} who identify the feature with a dusty clump
detected in CO and conclude, similarly, that it is located on the far side 
of the bulge. We will thus refer to the feature in question 
as the ``rear filament". However, an adjacent filament which extends 
from the nucleus (``+") to point ``C" is seen prominently in 
both emission and absorption, indicating that it is closer to us. 
This segment is part of a longer filament which passes
directly over the nucleus in line-of-sight projection. We refer to this
filament as the ``central filament".
Based on its elevated dust temperature 
(see Fig. \ref{fig4}e), it likely passes directly 
through the centre of the bulge where the interstellar radiation field 
due to evolved giant stars is strongest. 
Its overall position angle is $\sim50^\circ$, although it contains a
central ``jog" whereby it runs approximately east-west over the nucleus.

If the central filament is the manifestation of a bar, representing 
orbital motion under the influence of a bar-like gravitational potential
\citep[see, for example ][]{sell93},
we would expect to see some evidence of it in the distribution 
of bulge stars. Panel (c) of Fig. \ref{fig8} shows that this is indeed
the case---the bar structure is clearly visible in a map of stellar density.
The latter represents the number density of 
2MASS\footnote{Two Micron All Sky Survey} sources brighter than
$14^{\rm th}$ magnitude at $J$ band, estimated using a circular Gaussian kernel
whose width (FWHM) corresponds to 2 pixels (8 arcsec).  The position
angle of the stellar bar is approximately $45^\circ$, reasonably consistent
with that of the central filament.
We thus interpret the central dust structure as a bar, and explain its 
apparently double appearance in the column density map as a 
line-of-sight juxtaposition
with the ``rear filament" which is actually located further back.

Further elucidation of the nature of the central structure can be obtained
from a comparison with radial velocity information. Since the morphology of the
dust distribution closely resembles that of the ionized gas \citep{li09},
we make use of radial velocities of [NII], published previously by 
\citet{boule87}. Such a comparison is facilitated by panel (d) of Fig. 
\ref{fig8}, which represents a superposition of the [NII] radial velocities 
on the dust column density distribution.
The general trend of the radial velocities (red-shifted at the upper left
and blue-shifted at the lower right) is consistent with the general
rotational motion in the kinematic model presented by \citet{melch11},
involving an inner disk tilted at a position angle of $70^\circ$ and an 
inclination of $43^\circ$ to the plane of the sky. For the reasons discussed
above, we now believe that this disk is actually a compact spiral containing
a classical bar.

Fig. \ref{fig8}(d) also shows that the profile of radial velocity 
along the central filament is
quite distinct from that of the general rotational motion. It changes
abruptly from a red-shift to a blue-shift as the nucleus position is
traversed, at which point the velocity gradient is approximately east-west,
consistent with the local orientation of the jog in the filament.
This velocity structure, together with the apparent positional 
coincidence between the central filament and nucleus, would indicate motion 
either towards or away from the nucleus. However, we can eliminate the latter 
possibility because of the orientation---the lower right portion of the
filament must be pointed away from us since it experiences more optical
extinction then the upper left portion. Such an orientation is, in fact,
consistent with the inferred tilt of the nuclear spiral as a whole,
based on the \citet{melch11} kinematic model.
We thus conclude that the red- and blue-shifts indicate motion {\it towards\/}
the nucleus, i.e., that the filament
constitutes an accretion channel. Since the filament appears to
connect with the nuclear spiral at diametrically opposite points, it
seems likely that these spiral arms provide the source of the inflow.

We can estimate the inflow rate given
the column density of the central filament ($\sim2\times10^{20}$ H$_2$ masses
per cm$^2$), the radial velocity $\sim100$ km s$^{-1}$,
the inclination ($\sim47^\circ$ to the line of sight) and an estimated width
of 8 arcsec (60 pc). We thereby obtain a total flow rate (from both sides) of
$\sim0.05$ $M_\odot$ yr$^{-1}$. This value greatly
exceeds, by 8 orders of magnitude, the minimum accretion rate
(based on mass-energy equivalence) required to power the central X-ray source, 
the estimated luminosity of which is 
$\simeq1.2\times10^{36}$ erg s$^{-1}$ \citep{li09}. It is also exceeds, by
3 orders of magnitude, the Bondi accretion rate estimated by Li et al.
It is, however, close
to the inflow rate required for continuous replenishment of the hot corona
according to the evaporation model of Li et al.

The role of the bar as a conduit for gas flow inwards from the spiral arms 
to the nucleus in barred galaxies has been established observationally
\citep{knapen02} and has been the subject of a number of theoretical 
and modelling studies in which the inflow is attributed to
bar-induced torques \citep[see, for example,][]{gar09,quer16,hop11}. 
It is not uncommon for
nesting of bars to occur, whereby a subkiloparsec (secondary) nuclear bar
exists within a large-scale primary bar \citep{laine02,shlos02,woz15}. 
We suggest
that M31 represents an example of such nesting. Future dynamical modelling
will be needed to determine whether our inferred nuclear bar can
supply sufficient torque to drive the suggested inflow,
and to establish the dynamical relationship between the primary and 
secondary bars.

%%%%%%%%%%
\section*{Acknowledgements}
%%%%%%%%%%

We thank Anne-Laure Melchior for interesting discussions on this work.
We gratefully acknowledge the support of a consolidated grant (ST/K00926/1), 
from the UK Science and Technology Funding Council. This work was performed 
using the computational facilities of the Advanced Research Computing at 
Cardiff (ARCCA) Division, Cardiff University. This publication makes use of
data products from the Two Micron All Sky Survey, which is a joint project of
the University of Massachussetts and the Infrared Processing and Analysis
Center/California Institute of Technology, funded by the National Aeronautics
and Space Administration and the National Science Foundation.

%%%%%
\appendix
%%%%%

\section{Extension of {\sc ppmap}\ to enable $\beta$ estimation}

The original version of {\sc ppmap} \citep{mar15} treated the opacity index,
$\beta_{_{\rm D}}$, as constant. We have extended the algorithm by
treating $\beta_{_{\rm D}}$ as an additional state variable, thus
enabling the generation of 4D hypercubes of differential column density
as a function of RA, Dec, $T_{_{\rm D}}$, and $\beta_{_{\rm D}}$.
In doing so it is necessary to guard against the well-known degeneracy
between $T_{_{\rm D}}$ and $\beta_{_{\rm D}}$, whereby the presence of
measurement noise induces an anticorrelation between the estimates of those
quantities. Various approaches have been proposed for dealing with this
problem, most of which are oriented towards fitting the observed spectral 
energy distributions (SEDs) of models of multiple uniform clumps in a
given field. In the Hierarchical Bayesian approach \citep{kelly12,gal18} the
solution is constrained by the use of a prior distribution designed to
represent the statistics of the source population as a whole and whose
parameters are estimated adaptively from the observations. 
The algorithm then works towards finding solutions which
minimize the dispersion of the estimated parameter values, and appears to
work well provided one can characterise the source population 
by a single mode. However, extending the hierarchical Bayesian scheme to the 
more general imaging problem would present difficulties since it would
belong in the category of single-pixel SED fitting techniques requiring all
input images to be smoothed to the same resolution, thus losing an important
advantage of {\sc ppmap}. In addition, one would need to assume a single 
temperature and opacity index along each line of sight and this, in itself,
leads to difficulties in deducing dust properties \citep{juv13}.  The
{\sc ppmap} procedure is not subject to that restriction.

We can, however, achieve the main goal of suppressing
the $T_{_{\rm D}}$-$\beta_{_{\rm D}}$ degeneracy using a
non-hierarchical Bayesian scheme in which we specify
{\it a priori\/} the population statistics regarding $\beta_{_{\rm D}}$.
For this purpose we assume that the {\it a priori\/} probability distribution 
of possible $\beta_{_{\rm D}}$ values is described by a Gaussian function 
characterised by mean and standard deviation 
${\bar\beta}$ and $\sigma_\beta$, respectively. The estimation
procedure is then essentially the same as described in \citet{mar15},
except as noted below. As before,
the desired differential column density, represented by 
vector $\rho$, is obtained 
numerically by a stepwise integration of the differential equation:
\begin{equation}
\frac{\partial\rho}{\partial t} + \phi_1\rho = 0\,,
\end{equation}
where $\phi_1$ is the conditioning factor which contains information on
both the measurement model and the data, and $t$ is a progress variable which
increases from 0 to 1 during the solution procedure. The only modification
necessary for incorporating the prior information regarding $\beta_{_{\rm D}}$ 
is that the initial condition, $\rho_{(t=0)}=\eta$ is now replaced with:
\begin{equation}
{\rho_n}(t=0)= \eta\times\alpha\exp[-(\beta_n-{\bar\beta})^2/(2\sigma_\beta^2)]\end{equation}
in which the index $n$ refers to the $n$th cell in the state space
for which the local value of $\beta_{_{\rm D}}$ is equal to $\beta_n$, and 
$\alpha$ is a normalisation constant designed to ensure that the average
value of $\rho_n(t=0)$ over all $n$ is equal to $\eta$. The latter quantity
represents the {\it a priori\/} dilution, as defined by \citet{mar15},
for which we have assumed a value of 0.3 in our M31 analysis.

For the Gaussian prior on $\beta_{_{\rm D}}$ 
we have assumed ${\bar\beta}$ = 2.0
and $\sigma_\beta$ = 0.35. In regions where the signal to noise ratio is high, 
these parameters have essentially no effect on the solution, so that no
bias results \citep[see, for example,][]{gal18}, and this is true of both
the hierarchical and non-hierarchical Bayesian approaches. However, in regions
of low signal to noise, the algorithm will default to the prior itself so that,
in essence, we are taking $\beta_{_{\rm D}}$ to be 2.0 except where
the data compel us otherwise.

%%%%%
\section{Supplementary material}\label{SEC:Online}

We have made available online\footnote{http://www.astro.cf.ac.uk/research/PPMAP\_M31/} the set of {\sc ppmap}\ output files for M31. They consist of a
set of files in FITS format which include:
\begin{enumerate}
 \item 4D hypercube of differential column density, 
	with axes RA, Dec, dust temperature ($T_{_{\rm D}}$), and dust 
	opacity index ($\beta_{_{\rm D}}$). The sets of grid values 
	of $T_{_{\rm D}}$ and $\beta_{_{\rm D}}$ are specified in the FITS 
	header. 
 \item 4D hypercube of $1\sigma$ uncertainty values of
				differential column density
 \item 2D map of integrated line-of-sight column
	density in units of $10^{20}$ H$_2$ molecules cm$^{-2}$,
				truncated at the $5\sigma$ level.
 \item 	2D map of mean line-of-sight dust opacity 
				index, truncated at the $5\sigma$ level.
 \item  2D map of mean line-of-sight dust temperature,
				truncated at the $5\sigma$ column density level.
 \item 2D map of the standard deviation of dust
				temperature along the line of sight, truncated
				at the $5\sigma$ column density level.
 \item 2D map of the skewness of the dust temperature
                                distribution along the line of sight, truncated
				at the $5\sigma$ column density level.
 \item 2D map of the kurtosis of the dust temperature
                                distribution along the line of sight, truncated
				at the $5\sigma$ column density level.
\end{enumerate}
%%%%%
%%%%%%%%
\label{lastpage}
\end{document}